# RFID-Cloud Smart Cart System

Yerlan Berdaliyev, Alex Pappachen James
Department of Electrical and Electronic Engineering
School of Engineering, Nazarbayev University
Astana, Kazakhstan
apj@ieee.org

*Abstract*— The main purpose of this work is in reducing the queuing delays in major supermarkets or other shopping centers by means of an Electronic Smart Cart System which will introduce an intellectual approach to billing process through RFID technology. Smart Cart System is a cooperative performance of three separate systems: a website developed for the shopping market, electronic smart cart device and anti-theft RFID gates. This project focuses on developing the electronic smart cart device itself. It involves an embedded electronic hardware that consists of an OLED display, Arduino Mega 2560 board, a specifically designed PCB, a Wi-Fi module, 13.56 MHz HF RFID reader, a power supply and a shopping cart.

I. INTRODUCTION

*A. Problem Statement*

Frequently, people encounter a problem of spending too much of their time waiting in queues for billing their purchases in different shopping centers or supermarkets. Waiting in queues negatively affects human morale and may cause misunderstandings or conflict amongst people, for instance, when someone breaks the line and stands in front of other people [1]. The proposed project aims to eliminate this problem by introducing a novel alternative to traditional billing methods, speeding up the payment process.

*B. Barcodes*

The vast majority of modern supermarkets use barcode system to identify products and check-in customers waiting in queue [2]. Barcodes represent a series of vertical black lines of different thickness and separation distance which can be coded into data information [2]. Figure 1a shows the barcode drawn on product packaging. The barcode reader, shown in figure 1b, reads the data represented by barcodes. In modern supermarkets, this data involves a unique ID of each product [2]. All information describing a particular product such as full name, cost, weight, etc. are stored in primary software database, and this product is addressed by unique ID that is read from barcodes [2]. This way, one can easily get all information about a particular product just by scanning barcode that is printed in some area on product packaging. Hence, existing system of barcodes in supermarkets induces possibilities such as automatic calculation of total cost of purchased items, generating bills and item listings.

The main problem with the existing system of barcode billing is the fact that each product is scanned only one at a time so that total scanning time grows gradually when there are plenty of purchased products. The barcode scanner is limited by direct visual contact with barcodes [2]. Thus, it cannot scan barcodes that stay out of its vision.

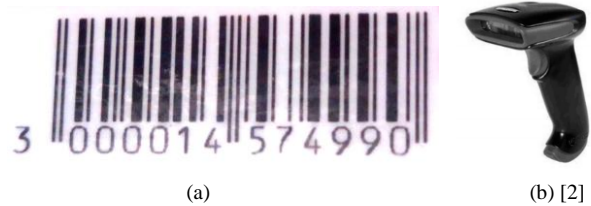

(a)       (b) [2]

Fig. 1. Barcode (a) and barcode reader (b) [2]

Therefore, people still tend to line up in queues in front of cashier's desk due to the inability of cashier's billing speed to catch up with the flow of new customers willing to check-out.

*C. Solution*

One measure to reduce the waiting time of customers is to introduce an intelligent billing system using electronic Smart Cart as an alternative to existing barcode system. Smart cart shown in Figure 2 allows a customer to manually perform billing without relying on cashier by means of swiping the RFID tags over RFID reader. Unlike barcode system, smart cart does not need any visual contact with barcodes which may get distorted in real life situations. All data about purchased products and user account data are stored in a cloud database in the Internet. Then, smart cart shows this information to customers on its display. A customer can delete an item from the list whenever he or she wishes. If the customer decides to finish purchasing, just a single button press is required to upload all purchased product data and their total cost to cloud database. Once all payment data is uploaded to the web, total cost is withdrawn from the registered account cash of the customer. All purchased products are deleted from the cloud database and the customer can freely pass the anti-theft gate with the purchased products.

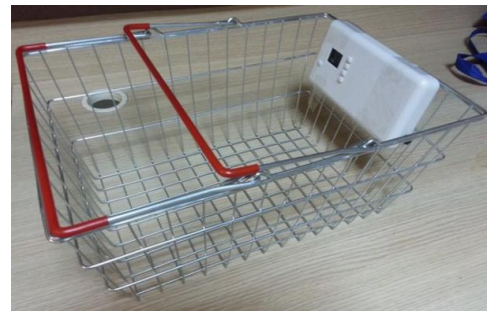

Fig. 2. The developed smart shopping cart

*D. Benefits*

In the community perspective, this project grants obvious benefits as it has the potential to decrease significantly the queuing time of customers and save much of precious time of every individual shopper. For instance, according to research [3] of British researchers, average queuing time in UK stores is 5 minutes 54 seconds, which is sufficient to miss most of the important activities such as airplane departure. On the other hand, in the market owners' and stakeholders' perspective, this system will be beneficial regarding attracting more customers, since their market will provide fast service and save shoppers time. Moreover, the new shopping experience and emerging technology attract more people, especially the young generation.

## II. BACKGROUND AND LITERATURE REVIEW

### A. RFID technology

One of the major technological advancements of recent years called RFID or Radio Frequency Identification made it possible to implement this project. RFID technology uses wireless propagation of electromagnetic wave signals over a certain frequency spectrum shown in figure 3 [4]. Generally, there are two types of RFID tags – passive and active [4]. Passive RFID tags shown in figure 4 do not have any power source inside their circuit [4]. Hence, they absorb enough electrical power for transmitting signals by harvesting RF power of receiver signals from their antenna that gets some energy from RFID reader or other sources [4]. The size of passive RFID tags can be tiny so that they can be attached to market products [4]. Passive tags are a major concern of this work.

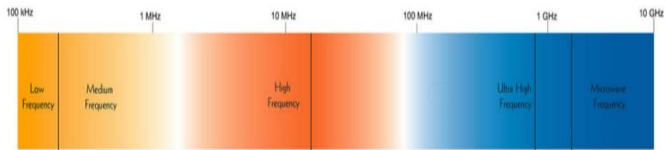

Fig. 3. The frequency spectrum of RF signals [4].

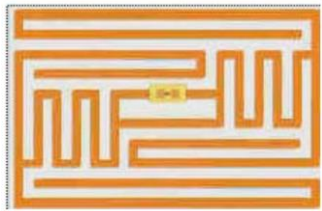

Fig. 4. An example of passive RFID tag/label used by Wal-Mart [4].

For instance, RF signals transmitted by RFID tags operating in UHF and MW frequencies hardly penetrate water containing bodies, thus, they work poorly near human tissue [5]. However, the reading distance and data rate of these tags are the largest [5]. On the other hand, LF and HF signals easily penetrate water environments, but RFID tags transmitting in these frequencies have lowest reading ranges and much less data rates than UHF and MW systems [5]. However, 13.56 MHz HF RFID tags have pretty acceptable data rates and reading distances and their signals easily penetrate humid environments [5]. Moreover, this frequency band gained high popularity and standardized in many countries [5]. There is a wide variety of commercial RFID tags working in 13.56 MHz frequency. Therefore, 13.56 MHz RFID tags and reader were chosen in this paper. Differences between the RFID technology and the existing system of barcode scanning are presented in table 1 below.

TABLE I.     COMPARISON OF BARCODE AND RFID TECHNOLOGIES

| Barcode scanners | RFID readers |
| --- | --- |
| Require close visual contact with the bar-coded print. | Do not require visual contact, but RFID tags must be brought to appropriate distance |
| Bar-coded prints may get distorted in wet or hot conditions or may get scratches. | RFID tags are much less affected by harsh environmental conditions. |
| Bar-code scanners are more expensive | RFID readers and tags are very cheap and massively produced |
| This technology is passing away | As time passes, more markets are turning to RFID tags. |

### B. Wi-Fi wireless transmission

Another important technology used in Smart Cart system is called Wi-Fi, which is a wireless data transmission technology or WLAN (Wireless Local Area Network) based on IEEE 802.11 standards [6]. Wi-Fi modules transmit data signals in 2.4 GHz or 5 GHz, which allows for very high data rates [6]. Currently, Wi-Fi is the most popular protocol for wireless connection and getting access to the Internet. Today, Wi-Fi hotspots can be found in almost every major building in a civilized country. Moreover, in some cities, whole area inside city borders is covered with free Wi-Fi hotspots which grants citizens free wireless access to the Internet [6]. In this work, smart carts use Wi-Fi technology for exchange of information read by RFID reader with the online cloud database which allows some important data to be stored securely in the Internet. Smart carts do not need to exchange large data such as digital documents, files, movies, etc. and therefore, they do not occupy significant portion of available bandwidth. Therefore, the exchange of information between a smart cart and market Wi-Fi hotspot happens instantly.

### C. Arduino

As the main processing unit of the Smart Cart hardware or brain and heart of the smart cart, an Arduino platform with Atmel microcontroller is used. The Arduino is an open-source prototyping platform based on easy-to-use hardware and software [7]. With the help of Arduino board, people with sufficient skills and knowledge can process input from different sensors and output processed data into different actuators, motors, displays or any other electrically controlled device [7]. For this project, the Arduino Mega 2560 board (illustrated in figure 5) with Atmel Atmega2560 8-bit AVR microcontroller is chosen due to its robustness, the majority of I/O pins and fast clocking rate [7]. However, Arduino boards are only useful for prototyping and not for commercial product design. At the project stage of designing final commercial product, an industry standard microcontroller platform will be chosen, and original PCB layout will be designed.

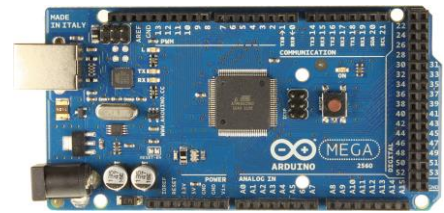

Fig. 5. Arduino Mega 2560 board [7].

### D. IBM Cloudant

In this project, almost all data related to market items and registered user accounts is stored in an online cloud database named Cloudant developed by IBM. As the company states, IBM Cloudant is a NoSQL database platform built for the cloud, which can be used as a fully-managed DBaaS (DataBase as a Service) running on public cloud platforms [8]. The key factor for choosing this database is the fact that it has RESTful API which allows easy access to the database from any device, including Arduino based platforms [8]. This allowed the Smart cart hardware to retrieve from or put data into the database by simply sending HTTP POST or PUT messages over TCP protocol. Moreover, Cloudant offers free account which contributed to the successful development of the project. Furthermore, Cloudant implemented secure connection based on SSL protocol which encrypts the transmitted data so that it will not be stolen or retrieved by unauthorized persons. All data in Cloudant database is stored in JSON format [8].

*E. Similar projects*

The idea of smart cart billing has gained a significant portion of interest among researchers which is proven by numerous existing articles on this topic. For instance, the article in reference [9] describes a similar project of Smart Cart and targets the same task of reducing queuing time in markets. In that article, each product that lies in the smart cart is scanned using RFID technology. The scanned information is sent to main server database through wireless channels using ZigBee protocol. However, the embedded system in that article is different from the proposed project by its hardware implementation and in few mechanisms. The proposed project is more sophisticated in technology, hardware, and software implementation. For example, the processing unit of embedded system in the current project is Atmega2560 of Arduino platform that is more robust, practical, low-power than the old 8051 platform used in the article [9]. Moreover, there is a plenty of advancements in this project such as a secure cloud database, fast data transmission over Wi-Fi protocol, original design of the casing, reliable and practical payment option that will be discussed later. There is another scientific research on the same field [10], in which the product data is read using barcode scanner that is good regarding cost, workload and already established infrastructure but requires a vision of the bar-code to scan. The bar-code scanner is built right into the smart cart. The amount of product is checked via weight scanner, and data is transmitted to the base station via ZigBee protocol, as in the first case. However, when the traditional bar-code scanner is used instead of RFID reader, it imposes several difficulties both to the customer and the market staff. Firstly, customers may face some issues when manually scanning bar-codes, since bar-codes may get visually distorted. Secondly, market staff will lose certain item tracking opportunities provided by RFID tags. Thirdly, bar-code readers are more expensive than RFID readers, which will significantly affect overall cost of the system when many smart carts are used. Moreover, there is the possibility of cheating of customers when they just put items into the smart cart without passing them through the bar-code scanner. Concerning this point, RFID technology is more secure regarding cheating because each RFID tag in the market can be tracked and unauthorized RFID tags (or unpaid products) just cannot be passed through the anti-theft security gates of supermarkets.

III. METHODOLOGY

In general, a market will have many Smart Carts that will serve a swarm of visitors. All these Smart Carts connect to the Internet through their Wi-Fi modules. Firstly, the market visitor who intends to use Smart Cart system registers and creates an account in the website provided by the market. After that, this person transfers some money into this account via internet banking or other sources. Therefore, the smart cart user will acquire some virtual cash in his/her market account. The value of this money and personal information along with purchase history will be stored in the Cloudant database. Market visitor can look at this information in the market website.

After creating the account and putting some cash into it, the visitor obtains a personal ID card provided by the market. Then, the visitor switches on the smart cart, swipes this ID card over it and authenticates himself in the Smart Cart System. After that, the visitor does shopping, chooses items with RFID tags and swipes these tags over the RFID reader antenna of the smart cart. Therefore, smart cart gets unique ID codes of each product lying inside it and transmits these codes to the Cloudant database, after which, retrieves all information related to these items. Smart cart hardware shows the information such as item names and costs on its built-in OLED screen. The customer looks into the display and can see the list of products lying inside the cart and all relevant information and total cost so that customer can make a better decision of what to purchase. Customer can easily remove an unwanted product out of the cart. However, it is expected that the customer will manually delete the item from the list by selecting it and pushing the "delete" button on the device. If the customer decides to finish purchasing and go billing, then he or she just presses the "pay" button on the smart cart which will send all purchased product information to the Cloudant database and will withdraw total cost from the personal cash of the user. These items will be deleted from the product database so that anti-theft security gates will not detect RFID tags attached to purchased items. However, if an item, which is not yet purchased, is passed across security gates, a security staff will approach the user and instruct how to use the Smart Cart system.

A clear illustration of integration and intercommunication of three systems is shown in the block diagram in figure 6. In this diagram, double-headed arrows show that communication goes in both directions and single-headed arrow represents one directional communication. For example, anti-theft gates only get information about the detected RFID tags from the Cloudant database while Smart Cart hardware reads and updates information in the database. Since the Cloudant database is already developed by IBM and commercially available, it is not regarded as a part of the Smart Cart system but intensively used.

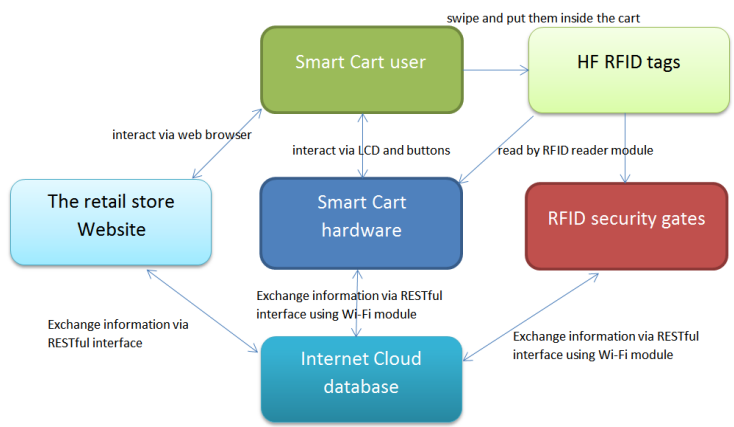

Fig. 6. Block diagram of the entire system.

Moreover, there are some basic steps that customer should do to accomplish the shopping process. A flowchart describing what a customer is expected to do is illustrated in figure 7.

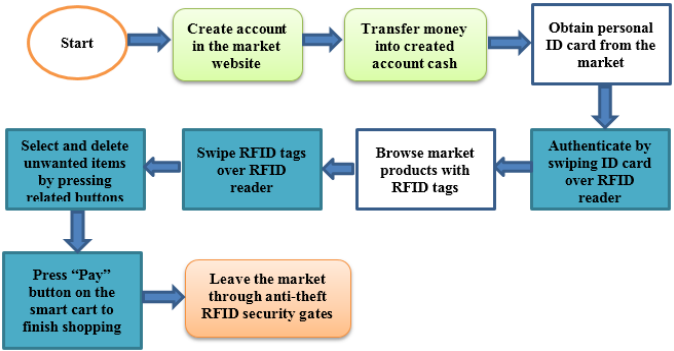

Fig. 7. Actions followed by the market visitor.

### A. The market website

The main functions of the market website are:

1. Provide convenient visual interface and user account
2. Communicate with IBM Cloudant and retrieve user data
3. Provide reliable money transfer from user bank accounts or web-money sources.
4. Show actual money stored in the user account and provide purchase history where purchased items are recorded for a short (or long) period of time.

The development of the website is not performed during this project, but it is left for market owners willing to use this Smart Cart System. Therefore, supermarkets are free to design their own website which will perform functions shown above. This website will not directly communicate with the smart cart hardware. Instead, it will retrieve the user cash balance and purchase history from the Cloudant database by sending HTTP GET requests. The RESTful API working in the Cloudant server will send back the required data in JSON format, encapsulated in HTTP protocol. The block diagram shown in figure 8 below shows the communication procedure and preliminary design of the user interface of the website. In this block diagram, the user of the Smart Cart system is named "Yerlan Berdaliyev" which the name of the author of this project. The name of the market is labeled as "market1".

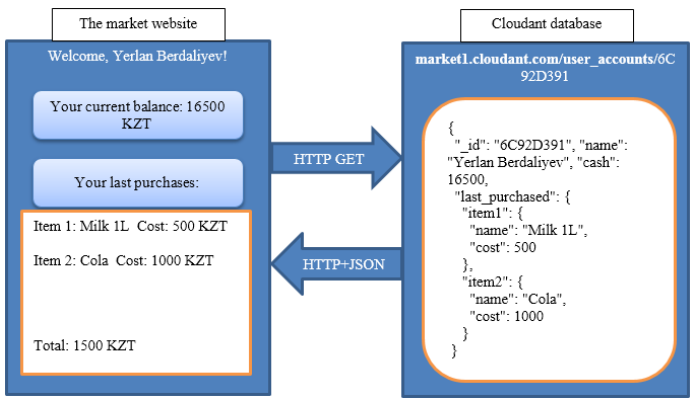

Fig. 8. Actions followed by the market visitor.

It can be seen that all data in the IBM Cloudant database is stored in so called JSON format, which is a very convenient way of representing structured key-value pairs. Firstly, the website API sends the user ID (which is 6C92D391 in the illustration above) in JSON format making HTTP GET request to Cloudant website. Then, Cloudant RESTful API sends back the stored information in JSON format as a response to GET request. Therefore, all necessary shopping information is shown in the market website.

### B. Anti-theft RFID security gates

Whenever a new technology is introduced to the general public, it often the case when someone tries to bypass security measures and cheats. The Smart Cart system is very vulnerable to such cheating actions of some dishonest users. In order to remove this vulnerability and prevent cheating, anti-theft security gates are used in the Smart Cart system. These gates are commonly used in the libraries or book stores. For this project, the manufacturer and model of security gates along with the software that runs these devices are left for the consideration of the market owners again. However, there are some general requirements for the security gates: detect all 13.56MHz RFID tags passing through the gate; have appropriate distance between two sides of the gate to let visitors pass through them; have connection to a computer or microcontroller which sends all read data of RFID tags to the Cloudant server and determines whether they exist in the database; have appropriate alarm system which clearly notifies market security staff. An example of anti-theft security gate is illustrated in figure 9 below. This security gate is the property of Nazarbayev University library.

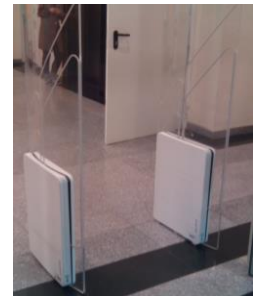

Fig. 9. Anti-theft security gate in Nazarbayev University library.

### C. Smart Cart hardware

Finally, in this section, the main focus of this project is discussed in depth. To clearly describe the Smart Cart hardware, this section was divided into several subsections.

*1) Hardware components*

The Smart Cart device involves a usual shopping cart with an electronic hardware device mounted on its wall. The electronic device consists of following components:

1. An Arduino Mega 2560 prototyping board, shown in figure 5 in chapter II

2. Espressif ESP-01 Wi-Fi module based on ESP8266 IC illustrated in figure 10 below. This module transmits and receives data in 2.4GHz with 802.11b/g/n protocol and requires only 3.3V power source. In order to successfully use this module with the Arduino board, it is required to upload the latest firmware into its IC. This firmware is provided by the Espressif company and newer versions are often released. For this project, the firmware version v0.51 was uploaded into the ESP-01 module. The communication between Arduino and the Wi-Fi module was done in serial port with 115200 bps baud rate. Furthermore, Arduino microcontroller talks to ESP8266 IC using "AT" commands developed by Espressif.

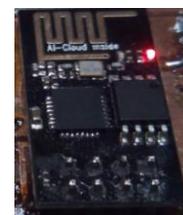

Fig. 10. ESP-01 Wi-Fi module.

3. Adafruit SSD1306 OLED display shown in figure 11 below. The resolution of this display is 128x64 and it requires 5V power supply. The communication is performed via I2C protocol. In Arduino 1.6.7 microcontroller programming software (which will be discussed later), the Adafruit GFX library provided by Adafruit GitHub page was used to program this display and output necessary information.

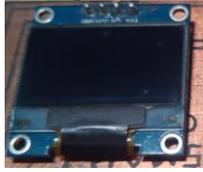

Fig. 11. Adafruit SSD1306 OLED display

4. 13.56 MHz (HF) RFID reader model Elechouse NFC Module V3 which is based on PN532 chip provided by NXP, shown in figure 12 below. Communication with this device is also run through I2C protocol. An Arduino library for communication with PN532 chip was provided by Elechouse on its GitHub page.

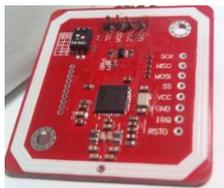

Fig. 12. Elechouse NFC v3 module

5. 9V battery and battery holder fitting PP3-size 9V battery.
6. Five standard 5mm x 5mm pushbuttons.
7. NXP BC556 PNP type bipolar junction transistor with hFe=125 current gain.
8. 500 Ohm resistor.
9. Power switch button.
10. Four jumper wires.

*2) Design and development of the PCB*

As shown above, Smart Cart hardware is composed of a number of electronic components. Initially, all these components were set up on a breadboard and tested. In order to bring a neat look into the Smart Cart, all these parts are assembled and mounted on a printed circuit board (PCB). The layout of this PCB is presented in figure 13 and it was drawn using computer software called Fritzing.

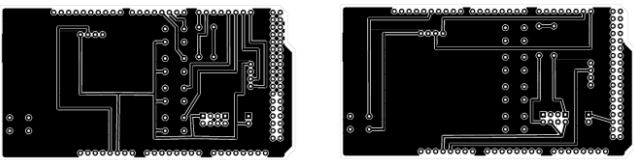

Fig. 13. Layout of the PCB. Top view (left) and bottom view (right)

This PCB was designed to be mounted on top of the Arduino Mega board, which significantly reduces the overall size of the device. The layout includes copper wire tracks connecting electrical components and holes on which they are set up and soldered. However, the majority of the PCB area is covered with copper fill. This was done for the purpose of speeding up the etching process. The layout of the PCB was printed using a laser printer on a white glossy paper. This layout was then translated onto the surface of a double-sided FR-4/copper board using electrical iron heater. This copper board with the design layout was then etched by placing into Iron(III) chloride ($FeCl_3$) solution. After some time, the PCB was ready for soldering components onto its surface. The final appearance of the PCB is presented later in the Results chapter.

*3) Programming the Arduino microcontroller*

The microcontroller of the Smart Cart device requires programming in order to perform given tasks. This is accomplished using Arduino IDE 1.6.7 software provided in the Arduino website. A popular programming language called C++ was used in the Arduino IDE. Every Arduino program includes two functions: setup() and loop(). In setup() function, all global variables are initialized and all external devices are turned on and initialized, since this function runs only one time during the startup of the Arduino. Moreover, some once performed tasks are also put into the setup() function. Therefore, all communications with the Wi-Fi, RFID reader, display modules are initialized and set up in the setup() function. On the other hand, loop() function contains commands or other nested functions that are repeated until Arduino is switched off or restarted. Therefore, all major tasks are placed in there. In loop function, firstly, Arduino checks the Wi-Fi connection, connection to the Cloudant server and connects if unconnected. Then, Arduino requires the user to authenticate and checks the ID card information by sending HTTP GET request to the Cloudant server and obtaining user information. After that, Arduino starts reading every RFID tag swiped over the RFID reader and exchange information with the Cloudant server. The constant run of the program can be interrupted when one of the five buttons are pressed. For instance, "up" or "down" buttons call buttonUp() or buttonDown() functions which allow user to select an item which can be deleted by pressing "delete" button. When user presses "pay" button, the buttonPay() function sends an HTTP PUT message to Cloudant server which updates user information on the database, subtracting total item price from the user cash and putting the purchased items in the purchase history. The final button is the "reset" button which restarts the device in case if failure occurs. There are plenty of other functions in the program code which perform minor tasks such as connecting to the Wi-Fi module, sending proper AT commands, finding required data in the response string etc.

*4) Design of the case for the Smart Cart hardware*

Like any other electronic device, an appropriate casing is needed to hold all components in one place, protect them from external harm and give the device a neat look. The casing for the Smart Cart hardware was designed using Dassault Systemes SolidWorks 2015 software. This design is presented in figures 14,15,16 below and it consists of two major parts and minor details. The dimensions of the case parts were determined in compliance with the dimensions of the Arduino Mega 2560 board, PCB and other related electronic components. The overall size of the casing was reduced to the minimum.

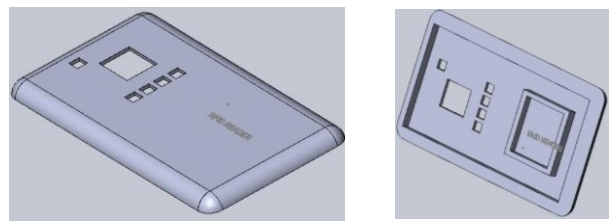

Fig. 14. The top part of the Smart Cart case. Isometric top view (left) and isometric bottom view (right).

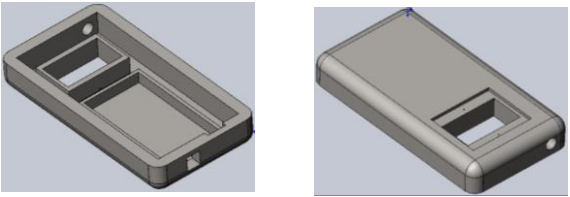

Fig. 15. The bottom part of the Smart Cart case. Isometric top view (left) and isometric bottom view (right).

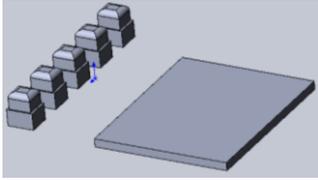

Fig. 16. The minor details of the Smart Cart case: 5 buttons and 9V battery cover for the bottom surface.

## IV. RESULTS AND DISCUSSION

### A. Printed Circuit Board

Photos of the PCB manufactured for this project are presented in figures 17&18 below.

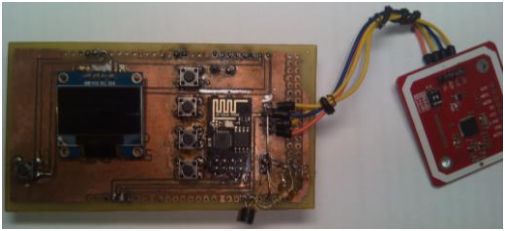

Fig. 17. The top view of the hardware assembly and PCB.

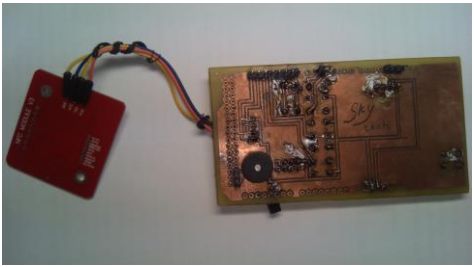

Fig. 18. The bottom view of the hardware assembly and PCB.

In two figures above, one can note that the copper wire tracks on FR-4 board and copper fill as well as electric components and modules soldered on it. Due to the lack of proper soldering equipment, some minor solder leaks were spread over the area of the PCB which significantly distorted the esthetic appearance of the PCB.

Moreover, one can note the improper placement of BJT transistor and a 500 Ohm resistor components on the edge of the PCB. Initially, BJT transistor and resistor components were not planned to be included in the hardware. However, after the process of manufacturing the PCB and soldering components on it, it was noted that buzzer produces very quiet noises when activated. Hence, it was decided to amplify the buzzer output using a BJT. Since the remake and redesign of the PCB takes a lot of time, it was decided to simply solder a BJT and a resistor on top of the PCB surface. In order to isolate vulnerable connections near the BJT and the resistor, a glue was poured on exposed conductors.

After soldering out all electronic components, the device was switched on and all modules were tested. All components worked properly, except the fact that the buzzer produced very quiet random noise during the device runtime. However, this noise is not noticeable and is negligible.

### B. Chassis

The manufactured result of the Smart Cart hardware with outer casing is illustrated in figures 20 and 21 below. The casing was 3D printed in SLT format using ABS material. This material is weaker and more brittle than plastic. However, it can successfully provide proper casing for the Smart Cart hardware and fix it in place. It can be observed that the casing lacks holes for screws to assemble top and bottom parts. Screws are avoided due to the weakness of ABS material. If some force is applied to the ABS wall with screws, whole wall structure can be collapsed. This is the main reason of avoiding screws. However, top and bottom parts can be fixed properly by sticking them together with some glue.

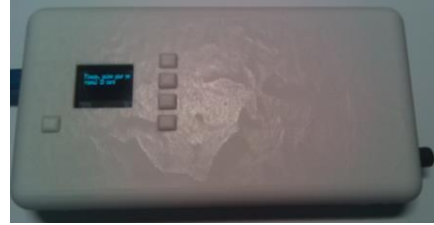

Fig. 19. The top view of the hardware assembly with outer casing.

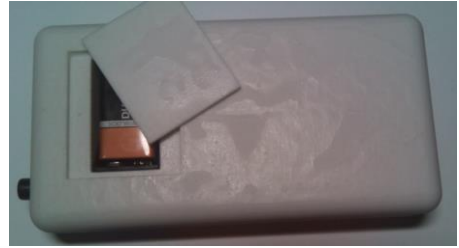

Fig. 20. The bottom view of the hardware assembly with outer casing.

### C. Results yielded by program code after uploading to Arduino

The hardware was tested with several RFID tags and the NU ID card of Yerlan Berdaliyev, who is the developer of this project. The performance results of the Arduino code uploaded into the Smart Cart microcontroller are clearly shown in the table 2 below. There are a lot of stages from powering on the hardware until the payment is completed. All these stages are described in the table.

TABLE II. TESTING OF THE PROGRAM CODE

| Stage # | Results output on the OLED display | Description of the stage |
|---|---|---|
| 1 | Hello! Starting up... | The Arduino microcontroller is currently running the setup() function which initializes all global variables and sets up communications with the modules. |
| 2 | Connecting WIFI | The Arduino microcontroller is communicating with the Wi-Fi module and instructs it to connect to a specific Wi-Fi access point. |
| 3 | 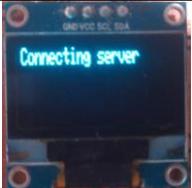 | The Arduino microcontroller is trying to connect to the IBM Cloudant server. It simply instructs the ESP-01 module to connect to IP 184.173.163.133 on port 80 with the TCP connection. |
| 4 | 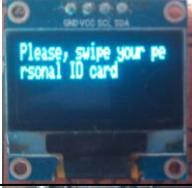 | The device is instructing user to swipe his/her ID card over the RFID reader antenna. After reading the ID card, the Smart Cart sends the ID code to Cloudant server and gets the user data in response. |
| 5 | 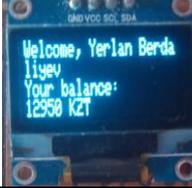 | In this stage, the device is showing some user data for 5 seconds, after which, it will step into the next stage. |
| 6 | 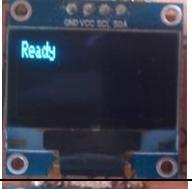 | This stage is showing that the Smart Cart device is ready to read RFID tags which are swiped over the RFID reader. |
| 7 | 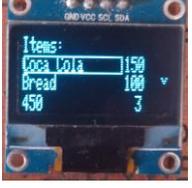 | In this stage, the device already recorded and obtained some data for 3 items. Item names are shown in the left and costs are presented in the right. Total cost of the items is shown in the bottom left corner and the total number of items is shown in the bottom right corner. First item is selected. Arrow down shows that there is a third item which is hidden. |
| 8 | 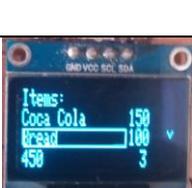 | This photo is similar to previous one. However, it clearly shows that the second item is selected using "down" button. |
| 9 | 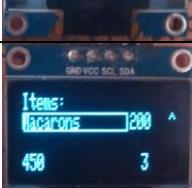 | This photo shows the third item which was not shown in the previous stages. Third item is also selected by pressing "down" button |
| 10 | 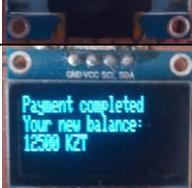 | Finally, "pay" button is pressed on the hardware which resulted in this photo. In this stage, the microcontroller sends the updated user information with the new balance and purchased items to the Cloudant server, after which, it will finish the session with the current user, delete all stored user data in the hardware memory and listen for the another user ID card. |

The total fabrication cost was approximately $100, excluding shopping cart. Smart Cart hardware can be easily attached on any shopping cart chosen by developers. This cost analysis is eligible only for the prototype device. However, if the Smart Cart system is ideally developed and brought into the commercial stage, the cost of one piece will significantly decrease since it will be mass produced.

The cost effectiveness is the extent of benefits regarding money spent. The Smart Cart system is cost effective in the long term because it needs infrastructure to be established. That is, RFID tags and readers are expected to get cheaper and widespread in near future [4]. Therefore, in the long term the value of benefits brought by Smart Cart system will outperform its production costs.

V. CONCLUSION

To conclude, this paper is about designing and developing a Smart Cart system that will simplify shopping experience of customers in supermarkets, reducing their waiting time and reducing human swarms in front of cashiers. RFID tracking is the key technology required for the implementation of this project. Smart Cart system involves effective communication between three separate systems: a website, Smart Cart electronic hardware and anti-theft RFID security gates. Main emphasis was made for the Smart Cart hardware while other two systems were left for the consideration of market owners. Therefore, a detailed design and development of the hardware was proposed and the practical model was successfully manufactured. This project proposes a novel approach in designing a Smart Cart system which is intended to reduce billing time in supermarkets. It is unique because it uses IBM Cloudant database for data storage and retrieval, one of the cheapest Wi-Fi modules commercially available and a cheap RFID reader. Moreover, an original design of the casing and PCB was proposed, which contribute to the uniqueness of this project. Finally, the working principle of the entire system is different from all other similar projects presented in the literature review chapter and it grants some flexibility in choice for market owners.